\documentclass[11pt,showpacs,aps,nofootinbib]{revtex4}
\usepackage{amsmath,amssymb,graphicx,epsfig,color,ulem,cancel,xcolor,datetime}
\def\comment#1{}

\def\nn{\nonumber}
\def\dst{\mathrm{dS}_{2}}

\def\dsf{\mathrm{dS}_{4}}

\def\rnd{\partial}
\def\x{\mathrm{x}}
\def\e{\rm\acute{e}}
\begin{document}
\title{Induced energy-momentum tensor of Dirac field in 2D de~Sitter~QED}
\author{Manizheh~Botshekananfard}\email{m.botshekananfard@gmail.com}
\author{Ehsan~Bavarsad}\email{bavarsad@kashanu.ac.ir}
\affiliation{Department of Physics, University of Kashan, 8731753153, Kashan, Iran}
\begin{abstract}
We compute the expectation value of the energy-momentum tensor in the in-vacuum state of the quantized Dirac field coupled to a uniform electric field
background on the Poincar$\e$ path of the two dimensional de~Sitter spacetime ($\dst$). The adiabatic regularization scheme is applied to remove the
ultraviolet divergencies from the expressions. We find, the off-diagonal components of the induced energy-momentum tensor vanishes and the absolute
values of the diagonal components are increasing functions of the electric field which decrease as the Dirac field mass increases.
We derive the trace anomaly of the induced energy-momentum tensor, which agrees precisely with the trace anomaly derived earlier in the literature.
We have discusses the backreaction of the induced energy-momentum tensor on the gravitational field.
\end{abstract}
\maketitle
\section{\label{sec:intro}introduction}
The theory of renormalization of the quantum electrodynamics (QED) in Minkowski spacetime is well established. Generally, in a curved spacetime and especially in the case of the de~Sitter spacetime (dS), there is no unique vacuum state to compute the expectation values of the quantities. Furthermore, the additional divergences are caused by the gravitational interactions. In order to establish the theory of renormalization in curved spacetime various methods have been developed; see \cite{Parker:2009uva,Birrell:1982ix} for introduction. Perhaps one of the most efficient methods to regularizing the expectation values of the quantities is the adiabatic subtraction method. This method is based on the approximation that the creation of a particle with
the frequency much larger than the energy scale determined by the curvature of the spacetime can be neglected. Hence, the adiabatic expansion is a power series of the spacetime curvature. Specifically, in the case of a Friedmann-Lemaitre-Robertson-Walker (FLRW) spacetime with a spatially flat metric the number time derivatives of the scale factor determines the adiabatic order. Adiabatic subtraction process starts from considering a Wentzel-Kramers-Brillouin (WKB) type solution for the under consideration mode equation and proceeds to construction of the Fock space of the state
vectors to compute the expectation values of the quantities; see \cite{Parker:2009uva,Birrell:1982ix} for details. Adiabatic regularization was introduced by Parker to obtain a finite expectation value of the particle number in an expanding universe \cite{Parker:1968mv}. After this original work, adiabatic regularization generalized and applied for various types of quantities and particles. Using adiabatic regularization scheme, the energy-momentum tensor
of a scalar field has been computed in spatially flat \cite{Fulling:1974zr,Parker:1974qw,
Bunch:1980vc,Habib:1999cs,LopezNacir:2007fdi,LopezNacir:2007wvc,Anderson:2013ila,Anderson:2013zia,Zhang:2019urk,Zhang:2019gtg} and closed \cite{Anderson:1987yt} FLRW type universes, anisotropic spacetimes \cite{Hu:1978zd,Nacir:2007dw}, and conformal coupling case \cite{Fulling:1974pu}.
The effective Lagrangian and the energy-momentum tensor of a scalar field in dS have been computed in Ref.~\cite{Dowker:1975tf} using the dimensional regularization method. Cosmological applications motivated authors of \cite{Mottola:1984ar,Markkanen:2016aes} to compute the energy-momentum tensor of
the created scalar particles in a $\dsf$. In \cite{Mottola:1984ar}, the finite energy-momentum tensor has been defined as difference of the expectation values in the in-vacuum and out-vacuum states; and the author found the decay of the effective cosmological constant due to the Hawking effect.
In \cite{Markkanen:2016aes}, a novel renormalization technique has been applied to compute the energy-momentum tensor, and the authors found that for
a massive noninteracting scalar field quantum corrections may lead to a superacceleration phase where the Hubble constant increase. In all the works \cite{Parker:1968mv,Fulling:1974zr,Parker:1974qw,Bunch:1980vc,Habib:1999cs,LopezNacir:2007fdi,
LopezNacir:2007wvc,Anderson:2013ila,Anderson:2013zia,Zhang:2019urk,Zhang:2019gtg,Anderson:1987yt,Hu:1978zd,Nacir:2007dw,Fulling:1974pu} the adiabatic regularization method developed and applied for a scalar field. The adiabatic expansion for the Dirac field has been introduced in \cite{Landete:2013axa} and further details provided in \cite{Landete:2013lpa}. The energy-momentum tensor of the Dirac field in a spatially flat FLRW spacetime has been regularized using the adiabatic subtraction prescription \cite{delRio:2014cha}; see also \cite{Ghosh:2015mva,Ghosh:2016epo,delRio:2017iib}. As proved in \cite{delRio:2014bpa} this method of the regularization gives the same result as the DeWitt-Schwinger point-splitting regularization, when applied to the energy-momentum tensor in a spatially flat FLRW universe.
\par
In this paper, we purpose to go beyond considering only a scalar or fermion field in a curved spacetime background, by adding an electromagnetic gauge
field to this picture. More specifically, we wish to consider a Dirac field coupled to a uniform electric field background in a dS. Creation of pairs in a strong electric field background in the flat spacetime is a well known nonperturbative phenomenon in quantum field theory which is referred to as the Schwinger effect \cite{Schwinger:1951nm,Sauter:1931zz,Heisenberg:1935qt}; see \cite{Gelis:2015kya,Dunne:2004nc} for introduction. Possibility of existing strong electromagnetic fields in the early universe \cite{Durrer:2013pga} motivates studying the Schwinger effect in the cosmological context \cite{Martin:2007bw}; see \cite{Kim:2019joy} for a recent review and references. Due to the cosmological event horizon dS emits particles \cite{Mottola:1984ar,Bousso:2001mw,Polyakov:2007mm,Polyakov:2009nq,Mahajan:2007qg,Haro:2008zz,Greenwood:2010mr,Anderson:2013ila,Anderson:2013zia,
Dabrowski:2014ica,Kim:2013cka,Tanhayi:2014ota} which is known as the Gibbons-Hawking radiation \cite{Hawking:1974sw,Gibbons:1977mu}.
In a dS, by using semiclassical approaches the Schwinger effect for scalar \cite{Garriga:1994bm,Garriga:1993fh,Frob:2014zka,Kobayashi:2014zza,
Bavarsad:2016cxh,Cai:2014qba,Kim:2016dmm,Kim:2008xv,Kim:2014iba,Geng:2017zad,Sharma:2017ivh,Hamil:2018rvu} and Dirac \cite{Villalba:1995za,Haouat:2012ik,
Havare:2002ra,Stahl:2015gaa,Hayashinaka:2016qqn,Stahl:2015cra,Haouat:2015uaa} fields has been investigated which may amplifies the Gibbons-Hawking radiation. Perturbative amplitude for production of the scalar \cite{Mihaela-Andreea:2014dda,Nicolaevici:2015qsa,Crucean:2015cga,Baloi:2019jui,
Crucean:2016sam} and Dirac \cite{Cotaescu:2010bn,Cosmin:2012zt,Crucean:2013sfg} pairs in the presence of an electric field on the dS has been studied and the authors found that the pair production is significant only when the expansion parameter is large compered to the mass of the field. Recently, the
effect of a magnetic field on the Schwinger effect in the $\dsf$ has been investigated \cite{Bavarsad:2017oyv}; see also \cite{Bavarsad:2018lvn,
Moradi:2009zz,Crucean:2016eiq,Nicolaevici:2016uzy,Crucean:2017pfg,Baloi:2018rcy,Baloi:2019wgd}, and it has been shown that the pair production is enhanced in the strong magnetic field regime. The Schwinger effect during inflation has been investigated using various approaches in Refs.~\cite{Stahl:2018idd,
Tangarife:2017rgl,Kitamoto:2018htg,Giovannini:2018qbq,Chua:2018dqh,Shakeri:2019mnt,Sobol:2018djj,Sobol:2019xls,Gorbar:2019fpj, Domcke:2019qmm}.
To provide further evidence for the ER=EPR conjecture, the holographic Schwinger effect in dS was studied in Ref.~\cite{Fischler:2014ama}.
\par
As noted in \cite{Frob:2014zka,Garriga:1994bm} the in-vacuum state of the quantum fields in dS has the Hadamard form, i.e., the ultraviolet divergencies
in the expectation values of the quantities in this state behaves similar to what is expected in the flat spacetime. Hence, it is preferred to study of
the renormalization theory in dS. In the pioneer work \cite{Frob:2014zka} the Schwinger effect and the in-vacuum expectation value of the induced current
of a scalar field in a $\dst$ has been investigated. The authors found that in the strong electric field regime, i.e., when the electric field is sufficiently large compered to the Ricci scalar curvature and the scalar field mass,
the current agrees with the semiclassical results \cite{Garriga:1994bm}.
A notable feature of the current was found in the weak electric field regime for a small enough scalar field mass compered to the Hubble constant, the current is inversely proportional to the electric field which leads to a phenomenon of infrared hyperconductivity. These remarkable results then motivate
to study the induced current for both scalar and fermions in various dimensions of the dS. The Schwinger effect and the adiabatic regularized induced current of the scalar field in a $\dsf$ has been investigated in \cite{Kobayashi:2014zza}, and particularly the phenomenon of infrared hyperconductivity
was reported. In order to compare the results, the vacuum expectation value of the induced current was renormalized using point-splitting method in \cite{Hayashinaka:2016dnt} which completely agrees with the work \cite{Kobayashi:2014zza}. In \cite{Geng:2017zad}, those results for the current have
been derived by using \textit{the uniform asymptotic approximation method}. The Schwinger effect in a general $D$-dimensional dS, and as an example of
odd dimensions, the induced current of the scalar field in the three dimensional dS have been investigated in \cite{Bavarsad:2016cxh} which a period of
the infrared hyperconductivity phenomenon was observed.
Furthermore, the authors of \cite{Bavarsad:2017oyv} consider a uniform magnetic field with conserved flux parallel to a constant energy density electric field in the $\dsf$ and find that the density of states are proportional to the magnetic field, consequently the scalar Schwinger effect is enhanced in
the sufficiently large magnetic field. They also show that in the sufficiently weak magnetic field the infrared hyperconductivity phenomenon occurs in
the induced current of the scalar field. For the case of the Dirac field the Schwinger effect and the induced current of the created pairs have been investigated in $\dst$ \cite{Stahl:2015gaa} and $\dsf$ \cite{Hayashinaka:2016qqn}. Contrary to the scalar field case, no infrared hyperconductivity phenomenon has been observed in the induced fermionic current. Analysing the induced current for both cases the scalar \cite{Kobayashi:2014zza,Hayashinaka:2016dnt} and Dirac \cite{Hayashinaka:2016qqn} fields in $\dsf$ reveals a period of the negative current where the current flows in opposite direction of the electric field background weaker than a threshold value depending on the field mass. The negative current
leads to the antiscreening effect which increases the magnitude of the electric field background. Obviously, the infrared hyperconductivity and the
negative current phenomena are contrary to physical intuition. In attempting to address these peculiarities, a new renormalization condition defined for
the scalar and Dirac induced currents in $\dsf$ which was named maximal subtraction \cite{Hayashinaka:2018amz}. The maximal subtraction scheme defines the asymptotic value of the renormalized vacuum expectation value equal to that of obtained from the semiclassical approaches. Following this scheme the infrared hyperconductivity period is removed, however a phase of the negative current would be present. Recently, in \cite{Banyeres:2018aax} all the contributions to the regularized induced current in $\dsf$ have been clarified. In all the works \cite{Kobayashi:2014zza,Bavarsad:2016cxh,Stahl:2015gaa,Hayashinaka:2016qqn} adiabatic subtraction scheme applied to regularize the induced current.
In de~Sitter spacetime with any dimension, the energy-momentum tensor of the semiclassical Schwinger scalar has been computed in the heavy scalar field \cite{Bavarsad:2016cxh} and the strong electric field \cite{Bavarsad:2017wbe} regimes; the authors found a decay of the Hubble constant due to the pair creation. The adiabatic regularized in-vacuum state expectation value of the trace of the induced energy-momentum tensor for a charged scalar field
coupled to a uniform electric field background in three \cite{Bavarsad:2018jpr} and four \cite{Bavarsad:2019jlg} dimensional dS has been studied.
In \cite{Bavarsad:2019jlg}, was shown that the gravitational backreaction effects of the trace of the induced energy-momentum tensor in the semiclassical regime leads to a decay of the Hubble constant, whereas in the infrared regime a superacceleration phenomenon occurs, i.e., the Hubble constant increases.
The method of adiabatic subtraction for regularization of both cases scalar and spinor QED in an expanding universe has been extensively studied in Refs.~\cite{Ferreiro:2018oxx,Ferreiro:2018qzr,Ferreiro:2018qdi,BarberoG.:2018qiv}.
In this paper our main goals are investigation of the induced energy-momentum tensor of the Dirac field in de~Sitter QED and further exploring validity
of the adiabatic subtraction regularization. Hence, we consider a massive charged Dirac field coupled to a uniform electric field background in a $\dst$.
To regularize the expectation value of the energy-momentum tensor in the in-vacuum state we adopt the second order adiabatic expansion of the appropriate counterterms.
\par
The paper is structured as follows. In Sec.~\ref{sec:model}, we briefly introduce our model.
In Sec.~\ref{sec:emt}, we compute the expectation value of the energy-momentum tensor and regularize it.
We examine the regularized energy-momentum tensor and discuses its consequences in Sec.~\ref{sec:result}.
Eventually, the conclusions are presented in Sec.~\ref{sec:concl}.
\section{\label{sec:model}the model}
We consider a massive Dirac field coupled to a uniform electric field with constant energy density on the Poincar$\e$ patch of $\dst$.
We assume the electric and gravitational fields as classical backgrounds which are not affected by the presence the Dirac field.
The half of $\dst$ can be represented as a spatially flat FLRW spacetime
\begin{align}\label{metric:flrw}
ds^{2}&=dt^{2}-e^{2Ht}d\x^{2}, & t&\in(-\infty,\infty), & \x&\in \mathbb{R},
\end{align}
where $t$ is the proper time and $H$ is the Hubble constant. By using the transformation
\begin{align}\label{conformal}
&\tau=-\frac{1}{H}e^{-Ht}, & \tau \in(-\infty,0)
\end{align}
the metric (\ref{conformal}) can be expressed in a manifestly conformally flat form
\begin{align}\label{metric:conf}
&ds^{2}=\Omega^{2}(\tau)\big(d\tau^{2}-d\x^{2}\big), & \Omega(\tau)=-\frac{1}{H\tau}.
\end{align}
In order to have a uniform electric field with a constant energy density in the metric (\ref{metric:conf}), we choose the electromagnetic vector
potential to be
\begin{equation}\label{potential}
A_{\mu}(\tau)=\frac{E}{H^{2}\tau}\delta_{\mu}^{1},
\end{equation}
where $E$ is a constant.
\par
The QED action for the Dirac field $\psi(x)$ with mass $m$ and charge $e$ which is coupled to the electromagnetic vector potential (\ref{potential}) in
the $\dst$ with metric (\ref{metric:conf}) can be written as
\begin{equation}\label{action}
S=\int d^{2}x\sqrt{-g}\bigg\{\frac{i}{2}\overline{\psi}(x)\Gamma^{\mu}\mathcal{D}_{\mu}\psi(x)
-\frac{i}{2}\overline{\psi}(x)\overleftarrow{\mathcal{D}_{\mu}}\Gamma^{\mu}\psi(x)-m\overline{\psi}(x)\psi(x)\bigg\},
\end{equation}
where $g$ is the determinant of the metric, and $\Gamma^{\mu}$ are the Dirac gamma matrices in the $\dst$. The metric tensor of the spacetime $g_{\mu\nu}$ can be expressed in terms of the Minkowski spacetime metric $\eta_{ab}$, by using the tetrad $e^{a}_{~\mu}$ coefficients
\begin{equation}\label{relation:tetrad}
g_{\mu\nu}(\tau)=e^{a}_{~\mu}(\tau)e^{b}_{~\nu}(\tau)\eta_{ab}.
\end{equation}
We choose the tetrad in the gauge
\begin{equation}\label{tetrad}
e^{a}_{~\mu}(\tau)=\Omega(\tau)\delta^{a}_{\mu},
\end{equation}
then, the Dirac gamma matrices in the Weyl representation are given by
\begin{align}\label{matrix:gamma}
&\Gamma^{0}=\Omega^{-1}(\tau)\sigma_{1}, & \Gamma^{1}=\Omega^{-1}(\tau)i\sigma_{2},
\end{align}
where $\sigma_{1},\sigma_{2},\sigma_{3}$ are Pauli matrices. In this representation the adjoint Dirac field $\overline{\psi}(x)$ is defined
\begin{equation}\label{adjoint}
\overline{\psi}(x)=\psi^{\dag}(x)\sigma_{1}.
\end{equation}
The gauge covariant derivatives can be expressed as
\begin{align}\label{derivative}
&\mathcal{D}_{\mu}=\rnd_{\mu}+B_{\mu}+ieA_{\mu},  &\overleftarrow{\mathcal{D}_{\mu}}=\overleftarrow{\rnd_{\mu}}-B_{\mu}-ieA_{\mu}
\end{align}
where the spin connection $B_{\mu}$ is given by
\begin{equation}\label{connection}
B_{\mu}=-\frac{\dot{\Omega}}{2\Omega}\sigma_{3}\delta^{1}_{\mu},
\end{equation}
where we use the over dot to denote the derivative with respect to the conformal time $\tau$.
\par
The Dirac equation follows from the action (\ref{action}),
\begin{equation}\label{diraceq}
\Big[i\Gamma^{\mu}\big(\rnd_{\mu}+B_{\mu}+ieA_{\mu}\big)-m\Big]\psi(x)=0.
\end{equation}
By using the fact that the Dirac Eq.~(\ref{diraceq}) is invariant under spatial translations, we can write $\psi(x)$ as decomposed two-component spinor
\begin{eqnarray}\label{spinor}
\psi(x)=\psi(\tau,\x)=e^{ik\x}
\left(
  \begin{array}{c}
    U_{1}(\tau) \\
    U_{2}(\tau) \\
  \end{array}
\right).
\end{eqnarray}
Substituting the decomposed form of the spinor (\ref{spinor}) into the Dirac Eq.~(\ref{diraceq}), and by using Eqs.~(\ref{potential}),
(\ref{matrix:gamma}) and (\ref{connection}) we arrive to a system of coupled deferential equations
\begin{eqnarray}
i\dot{U}_{1}(\tau)+q(\tau)U_{1}(\tau)+i\frac{\dot{\Omega}(\tau)}{2\Omega(\tau)}U_{1}(\tau)-m\Omega(\tau)U_{2}(\tau)=0, \nn \\
i\dot{U}_{2}(\tau)-q(\tau)U_{2}(\tau)+ \frac{i\dot{\Omega}(\tau)}{2\Omega(\tau)}U_{2}(\tau)-m\Omega(\tau)U_{1}(\tau)=0, \label{coupleqs}
\end{eqnarray}
where we have defined
\begin{equation}
q(\tau)=k+eA_{1}(\tau). \label{def:q}
\end{equation}
The system of coupled first order linear deferential equations (\ref{coupleqs}) can be converted into an equivalent two single second order linear
deferential equations
\begin{eqnarray}
\ddot{U}_{1}(\tau)+\Bigg(\omega_{k}^{2}(\tau)-iq(\tau)\bigg(\frac{\dot{q}(\tau)}{q(\tau)}-\frac{\dot{\Omega}(\tau)}{\Omega(\tau)}\bigg)
+\bigg(\frac{\ddot{\Omega}(\tau)}{2\Omega(\tau)}-\frac{3\dot{\Omega}^{2}(\tau)}{4\Omega^{2}(\tau)}\bigg)\Bigg)U_{1}(\tau) &=& 0, \nn \\
\ddot{U}_{2}(\tau)+\Bigg(\omega_{k}^{2}(\tau)+iq(\tau)\bigg(\frac{\dot{q}(\tau)}{q(\tau)}-\frac{\dot{\Omega}(\tau)}{\Omega(\tau)}\bigg)
+\bigg(\frac{\ddot{\Omega}(\tau)}{2\Omega(\tau)}-\frac{3\dot{\Omega}^{2}(\tau)}{4\Omega^{2}(\tau)}\bigg)\Bigg)U_{2}(\tau) &=& 0, \label{singleqs}
\end{eqnarray}
where the conformal time dependent frequency is given by
\begin{equation}
\omega(\tau)=+\sqrt{q^{2}(\tau)+m^{2}\Omega^{2}(\tau)}. \label{omega}
\end{equation}
The solutions of the Dirac Eqs.~(\ref{singleqs}) with the desired asymptotic forms at early times $\tau\rightarrow-\infty$ and late times $\tau\rightarrow0$, which describe the in and out-vacuum states respectively, have been obtained in Ref.~\cite{Stahl:2015gaa}. Since, we wish to compute
the expectation value of the energy-momentum tensor in the Hadamard in-vacuum state, we take only the solutions of Eqs.~(\ref{singleqs}) with the desired asymptotic behaviour at early times. Hence, the components of the normalized positive frequency mode spinor are given by \cite{Stahl:2015gaa}
\begin{eqnarray}
U_{1k}(\tau) &=& -i \mu e^{\frac{i\pi\kappa}{2}} W_{\kappa-\frac{1}{2},\gamma}\big(-2ip\big) \Theta(k)
+ e^{-\frac{i\pi\kappa}{2}} W_{-\kappa+\frac{1}{2},\gamma}\big(-2ip\big) \Theta(-k), \label{u1} \\
U_{2k}(\tau) &=& e^{\frac{i\pi\kappa}{2}} W_{\kappa+\frac{1}{2},\gamma}\big(-2ip\big) \Theta(k)
-i \mu e^{-\frac{i\pi\kappa}{2}} W_{-\kappa-\frac{1}{2},\gamma}\big(-2ip\big) \Theta(-k), \label{u2}
\end{eqnarray}
and, the components of the normalized negative frequency mode spinor are obtained \cite{Stahl:2015gaa}
\begin{eqnarray}
V_{1k}(\tau) &=& e^{\frac{i\pi\kappa}{2}} W_{-\kappa+\frac{1}{2},\gamma}\big(2ip\big) \Theta(k)
-i \mu e^{-\frac{i\pi\kappa}{2}} W_{\kappa-\frac{1}{2},\gamma}\big(2ip\big) \Theta(-k), \label{v1} \\
V_{2k}(\tau) &=& -i \mu e^{\frac{i\pi\kappa}{2}} W_{-\kappa-\frac{1}{2},\gamma}\big(2ip\big) \Theta(k)
+ e^{-\frac{i\pi\kappa}{2}} W_{\kappa+\frac{1}{2},\gamma}\big(2ip\big) \Theta(-k), \label{v2}
\end{eqnarray}
where $W$ is the Whittaker function \cite{NIST}, and the normalized step function has been defined as
\begin{equation}\label{step}
\Theta(k)=\sqrt{\frac{H}{2|k|}}\times\Bigg\{
\begin{array}{cc}
1 & \hspace{1cm} k>0, \\
0 & \hspace{1cm} k<0.
\end{array}
\end{equation}
We have defined the dimensionless variables
\begin{align}\label{variables}
&p =-|k|\tau, & \mu&=\frac{m}{H}, & \lambda&=\frac{eE}{H^{2}}, \nn \\
&r =\frac{k}{|k|}, & \kappa&=-i\lambda, & \gamma&=i\sqrt{\mu^{2}+\lambda^{2}}.
\end{align}
The normalized mode spinor components which are given by Eqs.~(\ref{u1})-(\ref{v2}), may be represented in the doublet form
\begin{align}\label{UandV}
&U_{k}(\tau)=\left(
       \begin{array}{c}
         U_{1k}(\tau) \\
         U_{2k}(\tau) \\
       \end{array}
     \right), &
&V_{k}(\tau)=\left(
       \begin{array}{c}
         V_{1k}(\tau) \\
         V_{2k}(\tau) \\
       \end{array}
     \right).
\end{align}
Then, it can be shown that these doublets satisfy the relation
\begin{equation}\label{relation}
U_{k}(\tau)U^{\dag}_{k}(\tau) + V_{k}(\tau)V^{\dag}_{k}(\tau)=\Omega^{-1}(\tau) \mathbb{I},
\end{equation}
where $\mathbb{I}$ denotes the $2\times2$ dimensional unit matrix.
\par
The set of complete orthonormal mode spinors (\ref{UandV}) determines the in-vacuum state $|0\rangle$ which is annihilated by acting the operators $a_{k}$ for a fermion and $b_{k}$ for an antifermion with the comoving momentum $k$, as
\begin{align}\label{vacuum}
&a_{k}|0\rangle=b_{k}|0\rangle=0, & \forall k.
\end{align}
Then, the Fock space of state vectors is built by successive acting the creation operators $a^{\dag}_{k}$ and $b^{\dag}_{k}$ for fermions and
antifermions, respectively. To quantize the Dirac field $\psi(x)$ we impose the anticommutation relations
\begin{equation}\label{anticommutation}
\{a_{k},a^{\dag}_{k'}\}=\{b_{k},b^{\dag}_{k'}\}=(2\pi)\delta(k-k').
\end{equation}
Then, the quantized Dirac field operator $\psi(x)$ can be written as
\begin{equation}\label{oprator:Dirac}
\psi(x)=\int_{-\infty}^{+\infty}\frac{dk}{(2\pi)}\Big[a_{k}U_{k}(\tau)e^{ik\x}+b^{\dag}_{k} V_{k}(\tau)e^{-ik\x}\Big].
\end{equation}
\section{\label{sec:emt}regularization of energy-momentum tensor}
The energy-momentum tensor of the Dirac field is defined by variation of the action~(\ref{action}) with respect to variation of the inverse metric
$\delta g^{\mu\nu}$, or more precisely
\begin{equation}\label{def:emt}
T_{\mu\nu}(x)=\frac{2}{\sqrt{|g|}}\frac{\delta S}{\delta g^{\mu\nu}}.
\end{equation}
After some algebra definition~(\ref{def:emt}) yields the expression for the energy-momentum tensor of the Dirac field, coupled to the uniform electric
field background in the $\dst$, as
\begin{equation}\label{expres:emt}
T_{\mu\nu}(x)=\frac{i}{4}\Big(\overline{\psi}\Gamma_{\mu}\mathcal{D}_{\nu}\psi+\overline{\psi}\Gamma_{\nu}\mathcal{D}_{\mu}\psi
-\overline{\psi}\overleftarrow{\mathcal{D}_{\nu}}\Gamma_{\mu}\psi-\overline{\psi}\overleftarrow{\mathcal{D}_{\mu}}\Gamma_{\nu}\psi\Big).
\end{equation}
\subsection{\label{sec:vev}In-vacuum state expectation values}
By considering the Dirac field $\psi(x)$ as the quantum operator which is given by Eq.~(\ref{oprator:Dirac}) and using the relations~(\ref{vacuum}) and (\ref{anticommutation}), we arrive at the following expressions for the in-vacuum state expectation values of the energy-momentum tensor components
\begin{eqnarray}\label{expres:00}
\langle 0 | T_{00} (x) | 0 \rangle &=& - \frac{H^{2}\Omega^{2}(\tau)}{2\pi}\Re\sum\limits_{r=\pm 1}\int_{0}^{\Lambda}\frac{dp}{p}
\bigg[p^{2}-\mu^{2}(p-\lambda r)e^{\pi\lambda r}\left| W_{i\lambda r-\frac{1}{2},\gamma}(2ip)\right|^{2} \nn \\
&+&i\mu^{2}e^{-\pi\lambda r} W_{-i\lambda r-\frac{1}{2},\gamma}(2ip)W_{i\lambda r+\frac{1}{2},\gamma}(-2ip) \bigg],
\end{eqnarray}
where $\Re$ denotes the real part, for the spacelike component
\begin{equation}\label{expres:11}
\langle 0 | T_{11} (x) | 0 \rangle = - \frac{H^{2}\Omega^{2}(\tau)}{2\pi}\sum\limits_{r=\pm 1}\int_{0}^{\Lambda}\frac{dp}{p}
\bigg[p^{2}-\mu^{2}(p-\lambda r)e^{\pi\lambda r}\left| W_{i\lambda r-\frac{1}{2},\gamma}(2ip)\right|^{2} \bigg],
\end{equation}
and the symmetric off-diagonal component obtained
\begin{equation}\label{vev:01}
\langle 0 | T_{01} (x) | 0 \rangle = \frac{H^{2}\Omega^{2}(\tau)}{2\pi}\lambda\Lambda.
\end{equation}
To regularize the ultraviolet divergencies in the momentum integrals of Eqs.~(\ref{expres:00})-(\ref{vev:01}), we have considered cutoff $K$ for the
comoving momentum $k$. Accordingly, cutoff $\Lambda=-K\tau$ is defined for the dimensionless physical momentum $p$; see definitions in Eq.~(\ref{variables}).
\par
To evaluate the integrals over momentum involving the Whittaker functions in Eqs.~(\ref{expres:00}) and (\ref{expres:11}), we follow the procedure that introduced in Refs.~\cite{Frob:2014zka,Kobayashi:2014zza} for computing the induced current of the scalar field in a $\dst$ and $\dsf$, respectively.
By using this method, the induced current of the Dirac field in a $\dst$ \cite{Stahl:2015gaa} and $\dsf$ \cite{Hayashinaka:2016qqn} has been computed.
The appropriate Mellin-Barnes integral representation \cite{NIST} of the Whittaker function is given by
\begin{eqnarray}
W_{\kappa,\gamma}(z) &=& e^{-\frac{z}{2}} \int_{-i\infty}^{+i\infty}\frac{ds}{2\pi i}
\frac{\Gamma\big(\frac{1}{2}+\gamma+s\big)\Gamma\big(\frac{1}{2}-\gamma+s\big)\Gamma\big(-\kappa-s\big)}
{\Gamma\big(\frac{1}{2}+\gamma-\kappa\big)\Gamma\big(\frac{1}{2}-\gamma-\kappa\big)} z^{-s}, \nn \\
\frac{1}{2}\pm\gamma-\kappa &\neq& 0,-1,-2,\cdots,  \label{Mellin}
\end{eqnarray}
which is vialed for arbitrary values of the phase of the argument in the domain $|\mathrm{ph}(z)|<3\pi/2$. The condition is that the contour of
integration must be arranged in a way that the chain of poles of $\Gamma(\frac{1}{2}+\gamma+s)\Gamma(\frac{1}{2}-\gamma+s)$ lies a part from the
chain of poles $\Gamma(-\kappa-s)$. By means of Eq.~(\ref{Mellin}) and the theorem of residues, we then obtain the final results
\begin{eqnarray}
\langle 0 | T_{00} (x) | 0 \rangle &=& \Omega^{2}(\tau)\frac{H^{2}}{4\pi} \bigg[ -2\Lambda^{2}-2\mu^{2}\log(2\Lambda)-2\lambda^{2} -\mu^{2}-i\gamma\sinh(2\pi\lambda)\csc(2\pi\gamma) \nn \\
&+&i\lambda+2\gamma\lambda\sinh(2\pi\lambda)\csc(2\pi\gamma)+\mu^{2}\Big(1+i\sinh(2\pi\lambda)\csc(2\pi\gamma)\Big) \nn \\
&\times&\psi\big(i\lambda+\gamma\big)
+\mu^{2}\Big(1-i\sinh(2\pi\lambda)\csc(2\pi\gamma)\Big)\psi\big(i\lambda-\gamma\big) \bigg], \label{vev:00}
\end{eqnarray}
and
\begin{eqnarray}
\langle 0 | T_{11} (x) | 0 \rangle &=& \Omega^{2}(\tau)\frac{H^{2}}{4\pi} \bigg[ -2\Lambda^{2}+2\mu^{2}\log(2\Lambda)-2\lambda^{2} -\mu^{2}+i\gamma\sinh(2\pi\lambda)\csc(2\pi\gamma) \nn \\
&-&i\lambda+2\gamma\lambda\sinh(2\pi\lambda)\csc(2\pi\gamma)-\mu^{2}\Big(1+i\sinh(2\pi\lambda)\csc(2\pi\gamma)\Big) \nn \\
&\times&\psi\big(i\lambda+\gamma\big)
-\mu^{2}\Big(1-i\sinh(2\pi\lambda)\csc(2\pi\gamma)\Big)\psi\big(i\lambda-\gamma\big) \bigg], \label{vev:11}
\end{eqnarray}
where $\log$ denotes the natural logarithm function, and the digamma function $\psi$ which is defined by the first derivative of the logarithm of the
Gamma function; see, e.g., \cite{NIST}.
\subsection{\label{sec:adi}Adiabatic regularization}
To remove the ultraviolet divergencies from the expectation values given by Eqs.~(\ref{vev:01}), (\ref{vev:00}) and (\ref{vev:11}) we apply adiabatic
subtraction scheme. This method has been used to regularize the induced fermionic current in $\dst$ \cite{Stahl:2015gaa} and $\dsf$ \cite{Hayashinaka:2016qqn}. To find the appropriate counterterms, we perform the same procedure as used in \cite{Stahl:2015gaa}. As usual in the
literature, we assume that the electromagnetic vector potential and energy-momentum tensor in the two dimensional spacetime to be of adiabatic order zero and two, respectively. We begin by considering a WKB type solution for Dirac equations (\ref{singleqs}) as
\begin{equation}\label{WKB}
\mathcal{U}_{a}(\tau)=N_{a} \exp\Big[ -i\int^{\tau}\Big(X_{a}(\tau')+iY_{a}(\tau')\Big)d\tau' \Big],
\end{equation}
where $a=1,2$ is the spinor index, $N_{a}$ is a normalization coefficient and the real functions $X_{a},Y_{a}$ are required to satisfy equations
\begin{eqnarray}
Y_{a}(\tau) &=& (-1)^{a} \frac{q(\tau)}{2X_{a}(\tau)}\Big(\frac{\dot{q}(\tau)}{q(\tau)}-\frac{\dot{\Omega}(\tau)}{\Omega(\tau)}\Big)
-\frac{1}{2}\frac{d}{d\tau}\log\Big(X_{a}(\tau)\Big), \label{Ya} \\
X_{a}^{2}(\tau) &=& Y_{a}^{2}(\tau)+\dot{Y}_{a}(\tau)+\omega^{2}(\tau)+\frac{\ddot{\Omega}(\tau)}{2\Omega(\tau)}
-\frac{3\dot{\Omega}^{2}(\tau)}{4\Omega^{2}(\tau)}. \label{Xa}
\end{eqnarray}
Inserting the function $Y_{a}$ from Eq.~(\ref{Ya}) into Eq.~(\ref{WKB}), we find that $\mathcal{U}_{a}$ can be expressed as
\begin{equation}\label{Ua}
\mathcal{U}_{a}(\tau)= \frac{N_{a}}{\sqrt{X_{a}(\tau)}} \exp\bigg(\int^{\tau} \bigg[-iX_{a}(\tau')
+(-1)^{a} \frac{q(\tau')}{2X_{a}(\tau')}\bigg(\frac{\dot{q}(\tau')}{q(\tau')}-\frac{\dot{\Omega}(\tau')}{\Omega(\tau')}\bigg) \bigg]d\tau' \bigg).
\end{equation}
Then, substituting $\mathcal{U}_{a}$ from Eq.~(\ref{Ua}) into decoupled Dirac Eqs.~(\ref{singleqs}) yields an algebraic equation for the function $X_{a}$,
which exactly is
\begin{eqnarray}\label{exactX}
X_{a}^{2}(\tau)-\omega^{2}(\tau) &=& -\frac{1}{2X_{a}}\Big(\ddot{X}_{a}-(-1)^{a}\ddot{q}\Big)
+\frac{1}{4X_{a}^{2}}\Big(3\dot{X}_{a}^{2}+\dot{q}^{2}\Big)
-(-1)^{a}\frac{\dot{X}_{a}q}{X_{a}^{2}}\bigg(\frac{\dot{q}}{q}-\frac{\dot{\Omega}}{\Omega}\bigg) \nn \\
&-&\frac{q^{2}}{4X_{a}^{2}}\frac{\dot{\Omega}}{\Omega}\bigg(2\frac{\dot{q}}{q}-\frac{\dot{\Omega}}{\Omega}\bigg)
-(-1)^{a}\frac{q}{2X_{a}}\bigg(\frac{\dot{\Omega}}{\Omega}
\frac{\dot{q}}{q}-\frac{\dot{\Omega}^{2}}{\Omega^{2}}+\frac{\ddot{\Omega}}{\Omega}\bigg)
-\frac{3\dot{\Omega}^{2}}{4\Omega^{2}}+\frac{\ddot{\Omega}}{2\Omega}.
\end{eqnarray}
The remaining undetermined factor in the adiabatic mode spinor (\ref{WKB}) is the normalization constant $N_{a}$. We will show below that the counterterms are independent of $N_{a}$. We constraint the components of the adiabatic mode spinor (\ref{WKB}) to obey the normalization condition
\begin{equation}\label{normalization}
\big|\mathcal{U}_{1}(\tau)\big|^{2}+\big|\mathcal{U}_{2}(\tau)\big|^{2}=\Omega^{-1}(\tau).
\end{equation}
Following adiabatic regularization method, the set of counterterms are constructed from the expectation values of the energy-momentum tensor components
(\ref{expres:emt}) in the adiabatic vacuum sate which is described by the solutions (\ref{WKB}). Such expressions can be written as
\begin{equation}\label{adi:00}
\mathcal{T}_{00} = \int_{-K}^{+K}\frac{dk}{(2\pi)}\Bigg[
\frac{q(\tau)\Big(\big|\mathcal{U}_{1}\big|^{2}-\big|\mathcal{U}_{2}\big|^{2}\Big)-m\Omega(\tau)\Big(\mathcal{U}_{1}\mathcal{U}_{2}^{\ast}
+\mathcal{U}_{1}^{\ast}\mathcal{U}_{2}\Big)}{\big|\mathcal{U}_{1}\big|^{2}
+\big|\mathcal{U}_{2}\big|^{2}} \Bigg],
\end{equation}
and, for the spacelike component
\begin{equation}\label{adi:11}
\mathcal{T}_{11} = \int_{-K}^{+K}\frac{dk}{(2\pi)} q(\tau) \Bigg[
\frac{\big|\mathcal{U}_{1}\big|^{2}-\big|\mathcal{U}_{2}\big|^{2}}{\big|\mathcal{U}_{1}\big|^{2}
+\big|\mathcal{U}_{2}\big|^{2}} \Bigg],
\end{equation}
where $K$ is an ultraviolet momentum cutoff. The off-diagonal component finally is obtained
\begin{equation}\label{adi:01}
\mathcal{T}_{01} = -\frac{1}{2}\int_{-K}^{+K}\frac{dk}{(2\pi)}q(\tau)
=\frac{H^{2}\Omega^{2}(\tau)}{2\pi}\lambda\Lambda.
\end{equation}
Defining the ratio of the spinor components
\begin{equation}\label{ratio}
\mathcal{R}=\frac{\mathcal{U}_{2}}{\mathcal{U}_{1}},
\end{equation}
one can rewrite Eqs.~(\ref{adi:00}) and (\ref{adi:11}) as
\begin{eqnarray}
\mathcal{T}_{00} &=& \int_{-K}^{+K}\frac{dk}{(2\pi)}\bigg[
\frac{q(\tau)\Big(1-\big|\mathcal{R}\big|^{2}\Big)-m\Omega(\tau)\Big(\mathcal{R}
+\mathcal{R}^{\ast}\Big)}{1+\big|\mathcal{R}\big|^{2}} \bigg], \label{emta:00} \\
\mathcal{T}_{11} &=& \int_{-K}^{+K}\frac{dk}{(2\pi)}q(\tau)\bigg[
\frac{1-\big|\mathcal{R}\big|^{2}}{1+\big|\mathcal{R}\big|^{2}} \bigg]. \label{emta:11}
\end{eqnarray}
The expressions for $\mathcal{U}_{1}$ and $\mathcal{U}_{2}$ can be read from Eq.~(\ref{Ua}) with $a=1$ and $a=2$, respectively. Substituting these
expressions into the coupled Dirac Eqs.~(\ref{coupleqs}), we then obtain
\begin{equation}\label{mathcall:R}
\mathcal{R}=\frac{1}{\Omega m}\bigg[X_{1}+q-\frac{i\big(\dot{X}_{1}+\dot{q}\big)}{2X_{1}}
+\frac{i\dot{\Omega}\big(X_{1}+q \big)}{2X_{1}\Omega}\bigg],
\end{equation}
where $X_{1}$ satisfies Eq.~(\ref{Xa}) with $a=1$. By virtue of Eqs.~(\ref{adi:00})-(\ref{mathcall:R}) the counterterms are independent of the
normalization constants $N_{a}$, as mentioned above.
\par
We now return to Eq.~(\ref{exactX}), to find a solution for the function $X_{1}$. Obviously, terms appearing on the right side of Eq.~(\ref{exactX}) are
of two higher adiabatic order than those on the left side. By using the fact that the conformal time dependent frequency $\omega$ [see Eq.~(\ref{omega})]
is of zero adiabatic order, at the lowest adiabatic order the function $X_{a}$ is obtained
\begin{equation}\label{Xorder:z}
X_{a}^{(0)}(\tau)=\omega(\tau),
\end{equation}
where the superscripts denote the adiabatic order.
Then, up to the second order the adiabatic expansion of $X_{a}$ is constructed iteratively and is given by
\begin{eqnarray}\label{Xorder:t}
X_{a}^{(2)}(\tau) &=& \omega(\tau)-\frac{1}{4\omega^{2}}\Big(\ddot{\omega}-(-1)^{a}\ddot{q}\Big)
+\frac{1}{8\omega^{3}}\Big(3\dot{\omega}^{2}+\dot{q}^{2}\Big)
-(-1)^{a}\frac{\dot{\omega}q}{2\omega^{3}}\bigg(\frac{\dot{q}}{q}-\frac{\dot{\Omega}}{\Omega}\bigg) \nn \\
&-&\frac{q^{2}}{8\omega^{3}}\frac{\dot{\Omega}}{\Omega}\bigg(2\frac{\dot{q}}{q}-\frac{\dot{\Omega}}{\Omega}\bigg)
-(-1)^{a}\frac{q}{4\omega^{2}}\bigg(\frac{\dot{\Omega}}{\Omega}
\frac{\dot{q}}{q}-\frac{\dot{\Omega}^{2}}{\Omega^{2}}+\frac{\ddot{\Omega}}{\Omega}\bigg)
-\frac{1}{2\omega}\bigg(\frac{3\dot{\Omega}^{2}}{4\Omega^{2}}-\frac{\ddot{\Omega}}{2\Omega}\bigg).
\end{eqnarray}
Hence, the adiabatic expansions of $X_{1}$, up the order of zero and two are read from Eqs.~(\ref{Xorder:z}) and (\ref{Xorder:t}) with $a=1$,
respectively. Then, substituting the second adiabatic order expansion of the expression on the right side of Eq.~(\ref{mathcall:R}) into
Eqs.~(\ref{emta:00}) and (\ref{emta:11}), we finally obtain
\begin{eqnarray}
\mathcal{T}_{00}^{(2)} &=& \Omega^{2}(\tau) \frac{H^{2}}{2\pi} \bigg[-\Lambda^{2}-\mu^{2}\log\big(2\Lambda\big)
+\frac{1}{12}+\mu^{2}\log(\mu)-\frac{\mu^{2}}{2}-\lambda^{2}+\frac{\lambda^{2}}{6\mu^{2}}\bigg], \label{mathcall:T00} \\
\mathcal{T}_{11}^{(2)} &=& \Omega^{2}(\tau) \frac{H^{2}}{2\pi} \bigg[-\Lambda^{2}+\mu^{2}\log\big(2\Lambda\big)
-\frac{1}{12}-\mu^{2}\log(\mu)-\frac{\mu^{2}}{2}-\lambda^{2}-\frac{\lambda^{2}}{6\mu^{2}}\bigg]. \label{mathcall:T11}
\end{eqnarray}
Subtracting the second adiabatic order counterterms, given by Eqs.~(\ref{adi:01}), (\ref{mathcall:T00}) and (\ref{mathcall:T11}) from the corresponding original expressions for the in-vacuum expectation values which are given by Eqs.~(\ref{vev:01}), (\ref{vev:00}) and (\ref{vev:11}), leads to the regularized induced energy-momentum tensor
\begin{eqnarray}\label{T00}
T_{00} &=& \langle 0 | T_{00} (x) | 0 \rangle - \mathcal{T}_{00}^{(2)} \nn \\
&=& \Omega^{2}(\tau)\frac{H^{2}}{4\pi}\bigg[-\frac{1}{6}-\mu^{2}\log(\mu^{2})-\frac{\lambda^{2}}{3\mu^{2}}-i\gamma\sinh(2\pi\lambda)\csc(2\pi\gamma)
+i\lambda \nn \\
&+&2\gamma\lambda\sinh(2\pi\lambda)\csc(2\pi\gamma)+\mu^{2}\Big(1+i\sinh(2\pi\lambda)\csc(2\pi\gamma)\Big)\psi\big(i\lambda+\gamma\big) \nn \\
&+&\mu^{2}\Big(1-i\sinh(2\pi\lambda)\csc(2\pi\gamma)\Big)\psi\big(i\lambda-\gamma\big) \bigg],
\end{eqnarray}
and
\begin{eqnarray}\label{T11}
T_{11} &=& \langle 0 | T_{11} (x) | 0 \rangle - \mathcal{T}_{11}^{(2)} \nn \\
&=& \Omega^{2}(\tau)\frac{H^{2}}{4\pi}\bigg[\frac{1}{6}+\mu^{2}\log(\mu^{2})+\frac{\lambda^{2}}{3\mu^{2}}+i\gamma\sinh(2\pi\lambda)\csc(2\pi\gamma)
-i\lambda \nn \\
&+&2\gamma\lambda\sinh(2\pi\lambda)\csc(2\pi\gamma)-\mu^{2}\Big(1+i\sinh(2\pi\lambda)\csc(2\pi\gamma)\Big)\psi\big(i\lambda+\gamma\big) \nn \\
&-&\mu^{2}\Big(1-i\sinh(2\pi\lambda)\csc(2\pi\gamma)\Big)\psi\big(i\lambda-\gamma\big) \bigg].
\end{eqnarray}
The important result is that the off-diagonal component of the regularized induced energy-momentum tensor vanishes
\begin{equation}\label{T01}
T_{01}= \langle 0 | T_{01} (x) | 0 \rangle - \mathcal{T}_{01} = 0.
\end{equation}
\section{\label{sec:result}examination of the induced energy-momentum tensor}
\begin{figure}[t]\centering
\includegraphics[scale=0.75]{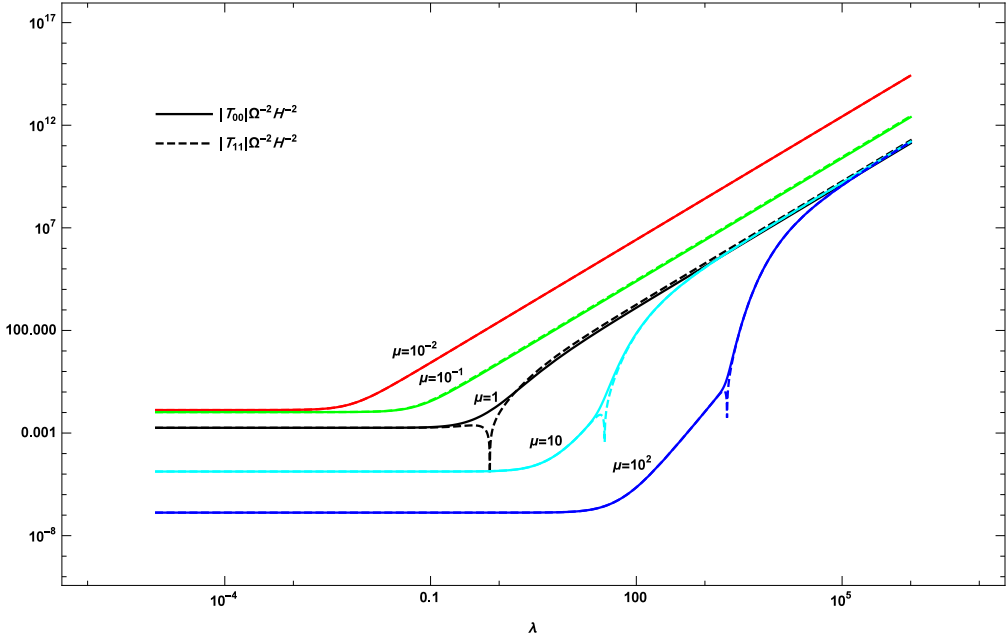}
\caption{The normalized magnitude of the induced energy-momentum tensor components, $|T_{00}|\Omega^{-2}H^{-2}$ in solid line, and
$|T_{11}|\Omega^{-2}H^{-2}$ in dashed line are plotted as functions of the normalized electric field $\lambda=eE/H^{2}$, for different values of the
normalized Dirac field mass $\mu=m/H$.} \label{fig:1}
\end{figure}
\begin{figure}[t]\centering
\includegraphics[scale=0.75]{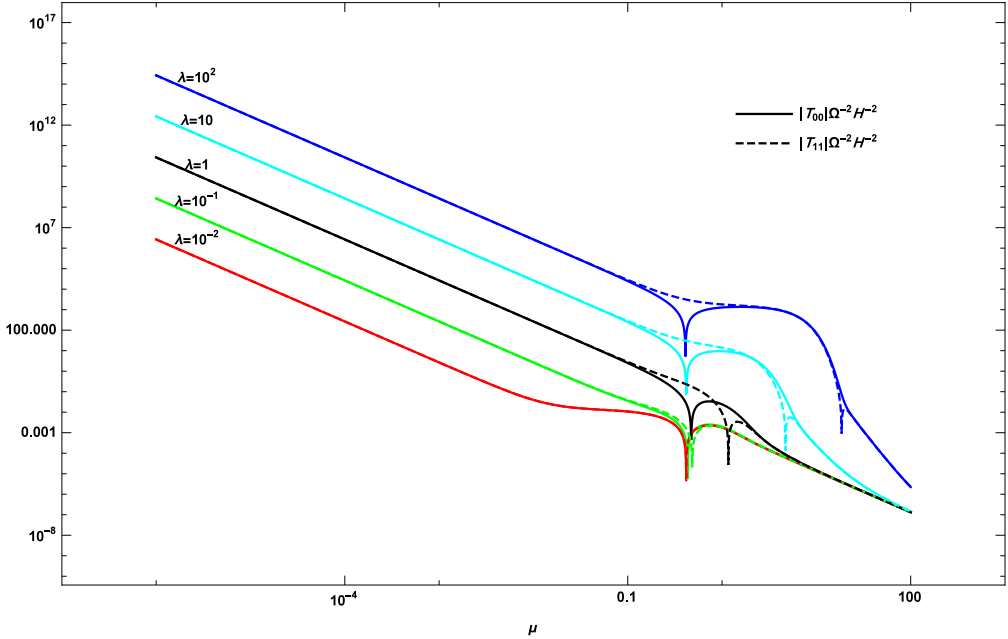}
\caption{The normalized magnitude of the induced energy-momentum tensor components, $|T_{00}|\Omega^{-2}H^{-2}$ in solid line, and
$|T_{11}|\Omega^{-2}H^{-2}$ in dashed line are plotted as functions of the normalized Dirac field mass $\mu=m/H$, for different values of the normalized
electric field $\lambda=eE/H^{2}$.} \label{fig:2}
\end{figure}

In this section we examine the regularized in-vacuum expectation value of the energy-momentum tensor which is given by Eqs.~(\ref{T00})-(\ref{T01}).
In Figs.~\ref{fig:1} and \ref{fig:2}, we have plotted the magnitude of $T_{00}$ (\ref{T00}) and $T_{11}$ (\ref{T11}) as functions of the electric field
and Dirac field mass, respectively. The qualitative features of these curves are the induced energy-momentum tensor is an increasing function of the
electric field which decrease with the Dirac field mass. Furthermore, the curves have discontinuity at which the sign of the expressions change.
\subsection{\label{sec:behav}Asymptotic behaviors}
We now analyze the induced energy-momentum tensor in the limiting values of the electric field and the Dirac filed mass, to find the explicit asymptotic
behaviors shown in Figs.~\ref{fig:1} and \ref{fig:2}.
\subsubsection{\label{sec:strong}{\bf Strong electric field}}
In the strong electric field regime, the asymptotic behavior of the induced energy-momentum tensor is obtained by expanding $T_{00}$ (\ref{T00}) and
$T_{11}$ (\ref{T11}) in a Taylor series about $\lambda=\infty$ with fixed $\mu$. We then obtain the leading order terms as
\begin{eqnarray}
T_{00} &\simeq& \Omega^{2}(\tau)\frac{H^{2}\lambda^{2}}{2\pi}\bigg(e^{-\frac{\pi\mu^{2}}{\lambda}}-\frac{1}{6\mu^{2}}\bigg), \label{strong00} \\
T_{11} &\simeq& \Omega^{2}(\tau)\frac{H^{2}\lambda^{2}}{2\pi}\bigg(e^{-\frac{\pi\mu^{2}}{\lambda}}+\frac{1}{6\mu^{2}}\bigg).  \label{strong11}
\end{eqnarray}
\subsubsection{\label{sec:weak}{\bf Weak electric field}}
In the weak electric field regime, the asymptotic behavior of the induced energy-momentum tensor is obtained by expanding $T_{00}$ (\ref{T00}) and
$T_{11}$ (\ref{T11}) in a Taylor series about $\lambda=0$. For the case of light Dirac field $\mu\ll 1$, we obtain the leading order terms
\begin{equation}\label{weak:light}
T_{00} = -T_{11} \simeq -\Omega^{2}(\tau)H^{2}\bigg( \frac{1}{24\pi}+\frac{\lambda^{2}}{12\pi\mu^{2}}
+\frac{\mu^{2}}{4\pi}\log\big(\mu^{2}\big)+\frac{\gamma_{\textrm{E}}\mu^{2}}{2\pi} \bigg),
\end{equation}
where $\gamma_{\textrm{E}}=0.577\cdots$ is the Euler's constant. And for the case of heavy Dirac field $\mu\gg 1$, the leading order terms are given by
\begin{eqnarray}
T_{00} &\simeq& \Omega^{2}(\tau)H^{2}\bigg(\frac{1}{240\pi\mu^{2}}
+4\lambda^{2}\mu e^{-2\pi\mu}-\frac{\lambda^{2}}{3\mu}e^{-2\pi\mu}\bigg), \label{weal:heavy00} \\
T_{11} &\simeq& \Omega^{2}(\tau)H^{2}\bigg(\frac{-1}{240\pi\mu^{2}}+\frac{\lambda^{2}}{3\mu}e^{-2\pi\mu}\bigg). \label{weal:heavy11}
\end{eqnarray}
\subsubsection{\label{sec:heavy}{\bf Heavy Dirac Field}}
For the case of heavy Dirac field, we can expand $T_{00}$ (\ref{T00}) and $T_{11}$ (\ref{T11}) in a Taylor series about $\mu=\infty$ with fixed $\lambda$.
We find that the expressions
\begin{eqnarray}
T_{00} &\simeq& \Omega^{2}(\tau)H^{2}\bigg(\frac{1}{240\pi\mu^{2}}+\frac{\lambda\mu}{\pi}e^{-2\pi\mu}\bigg),
\label{hevay00} \\
T_{11} &\simeq& \Omega^{2}(\tau)H^{2}\bigg(\frac{-1}{240\pi\mu^{2}}
+\frac{\lambda}{12\pi\mu}\big(1-4\lambda^{2}\big)e^{-2\pi\mu}\bigg), \label{heavy11}
\end{eqnarray}
well approximate the behavior of the diagonal components of the induced energy-momentum tensor as long as $\lambda\ll\mu$.
\subsection{\label{sec:trace}Trace of the induced energy-momentum tensor}
\begin{figure}[t]\centering
\includegraphics[scale=0.75]{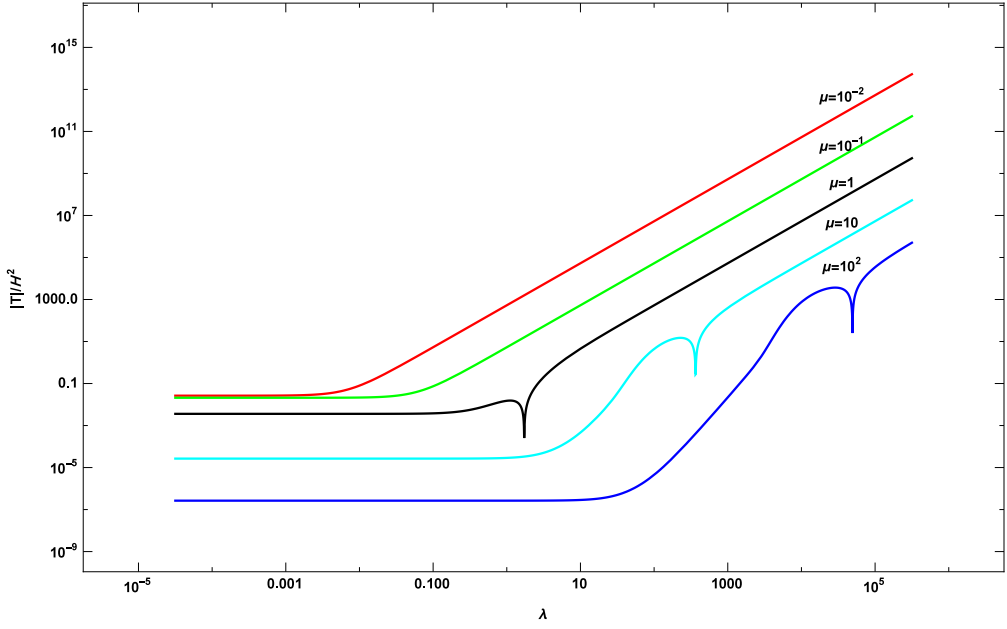}
\caption{The normalized magnitude of the trace of the induced energy-momentum tensor $|T|/H^{2}$ is plotted as function of the normalized electric field
$\lambda=eE/H^{2}$, for different values of the normalized Dirac field mass $\mu=m/H$.} \label{fig:3}
\end{figure}
\begin{figure}[t]\centering
\includegraphics[scale=0.75]{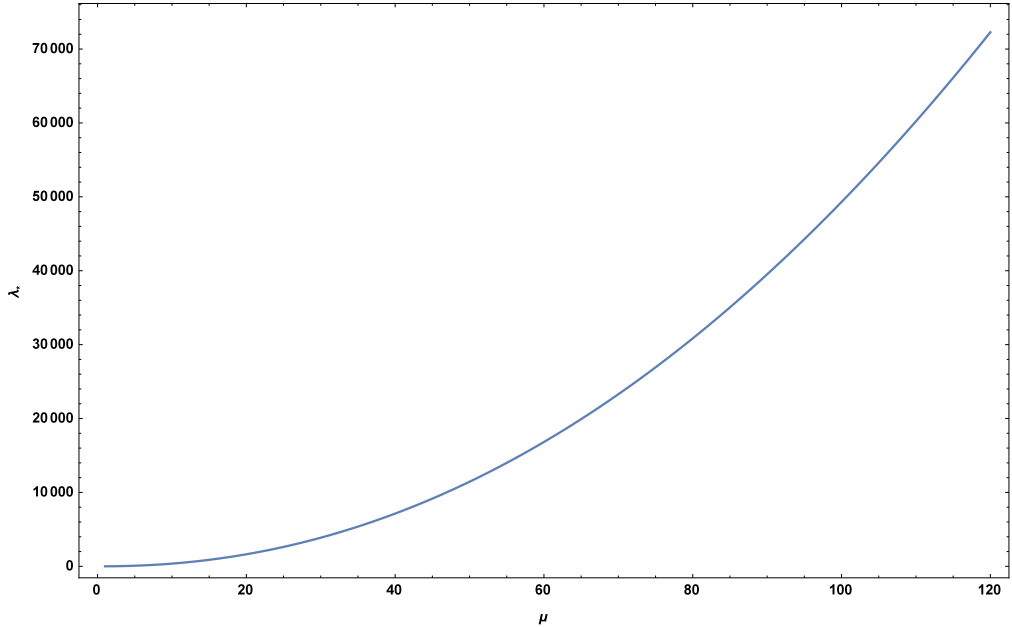}
\caption{The values of $\lambda_{\ast}$ are plotted against the normalized Dirac field mass $\mu$ at which the evaluation was made.} \label{fig:4}
\end{figure}

The trace of the induced energy-momentum tensor reads from Eqs.~(\ref{T00}) and (\ref{T11}),
\begin{eqnarray}\label{trace}
T &=& g^{\mu\nu}T_{\mu\nu} \nn \\
&=& \frac{H^{2}}{2\pi}\bigg[-\frac{1}{6}-\mu^{2}\log\big(\mu^{2}\big)-\frac{\lambda^{2}}{3\mu^{2}}-i\gamma\sinh(2\pi\lambda)\csc(2\pi\gamma)
+i\lambda \nn \\
&+&\mu^{2}\Big(1+i\sinh(2\pi\lambda)\csc(2\pi\gamma)\Big)\psi\big(i\lambda+\gamma\big)
+\mu^{2}\Big(1-i\sinh(2\pi\lambda)\csc(2\pi\gamma)\Big)\psi\big(i\lambda-\gamma\big) \bigg],
\end{eqnarray}
whose magnitude has been plotted in Fig.~\ref{fig:3}. This figure shows that the magnitude of the trace is an increasing function of the electric field
and decreases as the Dirac field mass increases. For the case of light Dirac field $\mu\lesssim 1$, the trace is negative $T<0$.
While, for the case of massive Dirac field $\mu\gtrsim 1$, the trace $T$ has a discontinuity at which its sign is changed. Let $\lambda_{\ast}$ be the
value of $\lambda$ at the discontinuity point of the trace $T$ whose precise value is evaluated by solving the equation
\begin{equation}\label{root}
T(\mu,\lambda_{\ast})=0.
\end{equation}
A numerical analysis illustrates that for $\lambda<\lambda_{\ast}$ the trace is positive $T>0$, whereas when $\lambda>\lambda_{\ast}$ the trace becomes
negative $T<0$. To find the behaviour of $\lambda_{\ast}$ against the Dirac field mass $\mu$ at which the evaluation was made, we present Fig.~\ref{fig:4}
which shows that $\lambda_{\ast}$ increases as $\mu$ increases.
\par
Before concluding, we would like to point out some consequences of the trace of the induced energy-momentum tensor, namely the trace anomaly and
gravitational backreaction.
\subsection{\label{sec:anomaly}{\bf Trace anomaly}}
The trace anomaly of the induced energy-momentum tensor is derived by taking the limit of $\lambda\rightarrow0$ and $\mu\rightarrow0$ in the expression
(\ref{trace}), which yields
\begin{equation}\label{anomaly}
T=-\frac{R}{24\pi},
\end{equation}
where $R=2H^{2}$ is the Ricci scalar curvature of $\dst$. This result agrees precisely with the well known trace anomaly \cite{Duff:1977ay} of a Dirac
field in the two dimensional de~Sitter spacetime.
\subsection{\label{sec:back}{\bf Gravitational backreaction}}
In Ref.~\cite{Stahl:2016geq}, the backreaction effect of the induced Dirac field current on the electromagnetic field background in $\dst$ has been
studied. The main conclusion was that the backreaction of the fermion pair creation does not amplifies the electric field background.
In this section we wish to study the backreaction of the induced energy-momentum tensor on the gravitational field background.
A natural way to do this is to use the Einstein gravitational field equations.
However, as is well known, there is no Einstein gravity in two dimensional spacetimes. Perhaps the most natural analogues of the Einstein gravitational
field equations in two dimensions is given by
\begin{equation}\label{Einstein}
R-\Lambda=8\pi GT,
\end{equation}
where $R$ is the Ricci scalar curvature, $\Lambda$ is the cosmological constant, $G$ is Newton's gravitational constant in two dimensions and $T$ is the
trace of the energy-momentum tensor; see, e.g.,\cite{Mann:1989gh,Mann:1991qp}.
\par
In a $\dst$ the cosmological constant vanishes, hence we set $\Lambda=0$ in Eq.~(\ref{Einstein}).
In natural units, Newton's gravitational constant $G$ should be dimensionless in two dimensional spacetimes.
Hence in a $\dst$ the Newton's gravitational constant, in terms of the Planck mass $M_{\rm P}$ and the Hubble constant $H$, can be written naturally as
$G=H^{2}/M_{\rm P}^2$. To explore the consequences of induced energy-momentum tensor, it is convent to define the time dependent Hubble parameter as
\begin{equation}\label{Hubble}
H(\tau)=\Omega^{-2}(\tau)\frac{d\Omega(\tau)}{d\tau}.
\end{equation}
The pair creation begins from early times, $\tau\rightarrow-\infty$, where the Ricci scalar curvature is given by $R=2H^{2}(\tau)$ and the induced
energy-momentum tensor of the created pairs is negligible compared to the Planck mass; and eventually, finishes at late times, $\tau\rightarrow0$, where
the Ricci scalar curvature approaches $R\approx0$, and the induced energy-momentum tensor of the created pairs is given by
Eqs.~(\ref{T00})-(\ref{T01}). Therefore, in this picture the time scale of the pair creation is of order of the Hubble time, i.e.,
$\delta\tau\approx H^{-1}$, and variations of the Ricci scalar curvature $\delta R$ and the trace of the induced energy-momentum tensor $\delta T$ are
given by
\begin{align}\label{variations}
& \delta R=-R, & \delta T=T,
\end{align}
where $T$ is given by Eq.~(\ref{trace}). Then, variation of both sides of Eq.~(\ref{Einstein}) leads to the evolution equation for the Hubble
parameter
\begin{equation}\label{evolution}
\dot{H}(\tau)=-\frac{2\pi H^{2}}{M_{\rm P}^2}T.
\end{equation}
Therefore, if $T>0$ then the Hubble constant decays. This condition holds for the case of heavy Dirac fields $\mu\gtrsim 1$ created by an electric filed
background with the strength $\lambda<\lambda_{\ast}$. Whereas, the Hubble constant amplifies and a superacceleration phenomenon with $\dot{H}>0$ occurs
in the condition that $T<0$. This is the case for light Dirac fields $\mu\ll1$ created by an unbounded electric filed strength or creation of heavy quanta
$\mu\gtrsim 1$ in an electric filed with the strength $\lambda>\lambda_{\ast}$.
\section{\label{sec:concl}conclusion}
In this paper we have computed the in-vacuum state expectation value of the energy-momentum tensor of a Dirac field in a uniform electric field background
on the Poincar$\e$ path of $\dst$. As expected the expectation values in the Hadamard in-vacuum state acquire quadratic ultraviolet divergences; see Eqs.~(\ref{vev:01}), (\ref{vev:00}) and (\ref{vev:11}). To obtain a finite result, we have applied the adiabatic regularization scheme.
Since the energy-momentum tensor is of seconde adiabatic order in two dimensions, we have constructed the set of the appropriate counterterms up to the second order of the adiabatic expansions; see Eqs.~(\ref{adi:01}), (\ref{mathcall:T00}) and (\ref{mathcall:T11}). Hence, we reach our goal of driving the induced energy-momentum tensor in Eqs.~(\ref{T00})-(\ref{T01}). We find that the off-diagonal components of the induced energy-momentum tensor vanish,
and the magnitude of the diagonal components are increasing functions of the electric field and decrease as the Dirac field mass increases; see Figs.~\ref{fig:1} and \ref{fig:2}. More precisely, these behaviours have been determined in Eqs.~(\ref{strong00})-(\ref{heavy11}).
\par
The trace of the induced energy-momentum tensor is given by Eq.~(\ref{trace}) and its magnitude is plotted in Fig.~\ref{fig:3}, that is an increasing
function of the electric field and decreases as the Dirac field mass increases. The analytic behaviour of the trace in the limiting values of the electric
field and mass can be read from Eqs.~(\ref{strong00})-(\ref{heavy11}). We find that for the case of heavy Dirac field $\mu\gtrsim1$, the trace has a discontinuity at which the sign of the trace changes. The strength of the electric field at the discontinuity point, which is denoted by $\lambda_{\ast}$, increases as the Dirac field mass increases; see Fig.~\ref{fig:4}.
We have shown that the induced energy-momentum tensor acquires the trace anomaly [see Eq.~(\ref{anomaly})] which precisely agrees with the trace anomaly
derived earlier in the literature for a Dirac field in $\dst$.
\par
In Sec.~\ref{sec:back}, we have briefly discussed the evolution of the Hubble constant caused by the induced energy-momentum tensor; see the final result
in Eq.~(\ref{evolution}). We find that creation of heavy Dirac fields $\mu\gtrsim 1$ in an electric filed whose strength is bounded as
$\lambda<\lambda_{\ast}$ leads to a decay of the Hubble constant due to the positive trace of the induced energy-momentum tensor.
Whereas, creation of light Dirac fields $\mu\ll1$ in an unbounded electric filed strength or heavy Dirac fields $\mu\gtrsim 1$ in an electric filed with
the strength of $\lambda>\lambda_{\ast}$ amplifies the Hubble constant and a superacceleration phenomenon with $\dot{H}>0$ occurs due to the negative
trace of the induced energy-momentum tensor.
\section*{\label{sec:acknow}ACKNOWLEDGMENTS}
E.~B.~is supported by the University of Kashan.

\end{document}